\documentclass[10pt,twocolumn]{article}
\usepackage{graphicx, subfigure, amsmath, algorithm, amssymb}
\usepackage{times}
\usepackage{fancyhdr}
\usepackage{tabularx}
\usepackage{hhline}
\usepackage{color}
\usepackage{alltt}
\usepackage{url}

\renewenvironment{thebibliography}[1]{%
    \begin{oldthebibliography}{#1}%
    \setlength{\parskip}{0ex}%
    \setlength{\itemsep}{0ex}%
}%
{%
    \end{oldthebibliography}%
}
\begin{document}
\title{The Case for Redundant Arrays of Internet Links (RAIL)}
\author{Athina Markopoulou, David Cheriton}
\date{}
\maketitle
\thispagestyle{plain}

\begin{abstract}
It is well-known that wide-area networks face today several
performance and reliability problems. In this work, we propose to
solve these problems by connecting two or more local-area networks
together via a Redundant Array of Internet Links (or {\em RAIL}) and
by proactively replicating each packet over these links. In that
sense, RAIL is for networks what RAID (Redundant Array of
Inexpensive Disks) was for disks. In this paper, we describe the
RAIL approach, present our prototype (called the {\em RAILedge}),
and evaluate its performance. First, we demonstrate that using
multiple Internet links significantly improves the end-to-end
performance in terms of network-level as well as application-level
metrics for Voice-over-IP and TCP. Second, we show that a delay
padding mechanism is needed to complement RAIL when there is
significant delay disparity between the paths. Third, we show that
two paths provide most of the benefit, if carefully managed.
Finally, we discuss a RAIL-network architecture, where RAILedges
make use of path redundancy, route control and application-specific
mechanisms, to improve WAN performance.

\end{abstract}

\section{Introduction}

The Internet gradually becomes the unified network infrastructure
for all our communication and business needs. Large enterprises, in
particular, rely increasingly on Internet-based Virtual Private
Networks (VPNs) that typically interconnect several, possibly
remote, sites via a wide-area network (WAN). Depending on the
company, the VPNs may have various uses, including carrying
Voice-over-IP (VoIP) to drive down the communication expenses,
sharing geographically distributed company resources, providing a
real-time service, etc.

However, it is well known that wide-area networks face today several
problems, including congestion, failure of various network elements
or protocol mis-configurations. These may result in periods of
degraded quality-of-service, or even lack of connectivity, perceived
by the end-user. To deal with these problems, several measures can
be taken at the end-points, at the edge, or inside the network.

One approach is to use redundant communication paths to improve
end-to-end performance\footnote{We consider reliability as an
extreme case of quality-of-service (QoS), because from a user's
perspective a ``failure'' has the same effect as several packets
lost in a row. At one extreme, packets may sporadically get dropped
or delayed - this is typically referred to as QoS problem. At the
other extreme, a failure may lead to an long-lasting loss of
connectivity - this is typically referred to as a reliability
problem. In the middle, several packets may get mistreated in a
short time period - which is also typically considered a QoS
problem. To cover the entire range of cases ,we often refer together
to quality-of-service and reliability, as ``performance''}. This
idea is not new. The Resilient Overlay Network (RON) architecture
\cite{ron} proposed that participating nodes maintain multiple paths
to each other, in order to preserve their connectivity in the face
of Internet failures. The more practical alternative to resilient
overlays, multi-homing \cite{akella1, akella2}, advocates that each
edge network connect to the Internet over multiple Internet Service
Providers (ISPs), in order to increase the probability of finding an
available path to any destination. Both approaches essentially
suggest to establish and intelligently use redundant communication
paths. Several vendors have already developed products along these
lines \cite{routescience, internap, fatpipe}. A significant body of
research has also investigated the performance of such approaches
and algorithms for monitoring, dynamic path switching and other
aspects \cite{ron,akella1, akella2,upenn1, upenn2,
chennee-cooexisting, dovrolis-infocom06, overlay-tcp, Kelly,
under-over, selfish-shenker}.

We too are looking at how to use control at the edge and utilize
redundant communication paths to improve end-to-end performance.
What we bring to the table is a mechanism for proactively leveraging
several paths at the same time. We propose to replicate and transmit
packets over several redundant independent paths, which are
carefully selected. The goal is to increase the probability that at
least one copy will be received correctly and on time. In other
words, we propose to combine a proactive replication over a set of
redundant links, with the traditional reactive dynamic switching
among (sets of) links.

Our approach is inspired by the Redundant Array of Inexpensive Disks
(RAID) \cite{raid}. The basic idea of RAID was to combine multiple
small, inexpensive disk drives into an array of disk drives which
yields better performance that of a Single Large Expensive Drive
(SLED), and appears to the computer as a single logical storage unit
or drive. Furthermore, disk arrays were made fault-tolerant by
redundantly storing information in various ways. Our approach is
analogous to ``disk mirroring'', or RAID-1, which duplicates all
content on a backup disk; so our approach would be called RAIL-1
according to RAID terminology.

Similarly to RAID, we propose to replicate packets over multiple,
relatively inexpensive, independent paths, i.e., create a {\em
Redundant Array of Internet Links (RAIL)}, which appears to the
application as a single ``superior'' link. To evaluate RAIL
performance, we have built a prototype called RAILedge. We show that
using RAIL yields better performance (both quality-of-service and
reliability) than using any of the underlying paths alone. In
addition, we evaluate the performance of applications, such as VoIP
and TCP, over RAIL and we seek to optimize relevant
application-level metrics. In particular, we propose an additional
mechanism, called {\em delay padding}, which complements RAIL when
there is a significant disparity between the underlying paths.

There are several issues that need to be investigated. How much is
the performance benefit from RAIL and how does it depend on the
characteristics of the underlying paths? What is the tradeoff
between performance benefit and the bandwidth cost of replicating
every packet over multiple connections? How does RAIL interact with
higher layers, such as TCP and VoIP applications? Does RAIL
introduce reordering? How should one choose the links that
constitute the RAIL, in a way that they complement each other and
optimize application performance? In this paper, we address these
questions.

With regards to the bandwidth cost, we argue that it is worthwhile
and that RAIL is a simple cost-efficient approach for achieving good
quality-of-service over redundant paths. The first argument is from
a cost point-of-view. As bandwidth gets cheaper and cheaper,
combining multiple inexpensive links becomes competitive to buying a
single, more expensive, private line. Furthermore, we show that two
paths are sufficient to get most of the benefit. In addition, the
cost of a connection is rather fixed than usage-based. Once one pays
the initial cost to get an additional connection to a second ISP
(which companies using multi-homing have already done), there is no
reason not to fully utilize it. The second argument is from a
performance point-of-view, which may be a strict requirement for
critical applications. RAIL-ing traffic over $n$ paths provides more
robustness to short term ``glitches'' than dynamic path switching
between the same $n$ paths. This is because there are limits in how
fast path switching mechanisms can (i) confidently detect glitches
and (ii) react to them without causing instability to the network.
For example, if a few VoIP packets are sporadically dropped, a path
switching system should probably not react to it, while RAIL can
still successfully deliver copies of the lost packets arriving from
the redundant paths.

Our findings can be summarized as follows.
\begin{itemize}
\item First, we demonstrate that proactively replicating packets
over a {\em Redundant Array of Internet Links (RAIL)} significantly
improves the end-to-end performance. We quantify the improvement in
terms of network-level as well as application-level metrics. In this
process, we use and derive analytical models for the performance of
VoIP-over-RAIL and TCP-over-RAIL. We also use a working prototype of
RAILedge.
\item
Second, we design and evaluate a {\em delay padding} mechanism to
complement RAIL when there is a significant delay disparity among
the underlying paths. This is useful both for VoIP (where it plays a
proxy-playout role) and for TCP (where it may remove re-ordering)
\item
Third, we show that two paths provide most of the benefit, while
additional paths bring decreasing benefits. The two preferred paths
should be carefully selected based on their quality,
similarity/disparity and correlation.
\end{itemize}

\begin{figure}[t]
\begin{center}
\centerline{\includegraphics[scale=0.37,angle=-90]{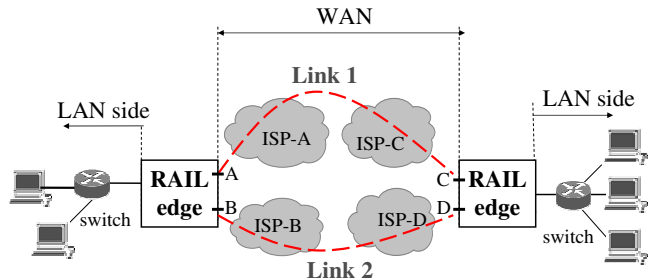}}
\end{center}
\vspace{-10pt} \caption{An example of a Redundant Array of Internet
Links (RAIL) connecting two remote sites.} \label{fig:topology}
\end{figure}

The structure of the rest of the paper is as follows. Section
\ref{sec:related} discuss related work. Section \ref{sec:system}
describes the RAILedge design, some implementation details and the
experimental setup. Section \ref{sec:evaluation} evaluates the
performance improvement brought by RAIL in terms of general
network-level metrics (subsection \ref{sec:network-level}), VoIP
quality (subsection \ref{sec:voip}) and TCP throughput (subsection
\ref{sec:tcp}); we also study the sensitivity to the characteristics
of the underlying paths. In this evaluation, we used analysis,
matlab simulation, actual packet traces collected over Internet
backbones, and testbed experiments. Section \ref{sec:discussion}
discusses the bigger picture, including possible extensions and open
questions. Section \ref{sec:conclusion} concludes the paper.

\section{Related Work}
\label{sec:related}

The use of redundancy is a well-known technique for improving system
reliability \cite{reliability-book}. In the networking context, a
common technique is to use redundant diverse paths in order to
improve the end-to-end performance. Multi-homing and routing
overlays both exploit path diversity, primarily to improve
availability in case of failures, and secondarily performance in
case of congestion in one of the two paths. Today, several vendors
provide services that combine multi-homing (i.e. the connection of
an edge network to several different ISPs) with additional control
capabilities at the edge (such as monitoring and dynamic ISP
switching, QoS mechanisms, compression) so as to optimize cost and
performance \cite{routescience,internap, fatpipe}. Overlay networks
provide additional control not only at the edge but also at
intermediate nodes \cite{ron}.

Several researchers are studying the performance of multi-homing and
overlay routing, and have proposed algorithms for monitoring and
path switching. The  pioneer Resilient Overlay Networks project is
described in \cite{ron}. Measurements-based performance evaluation
of multi-homing can be found in \cite{akella1, akella2}. The benefit
from path switching and the effect on application performance was
quantified in \cite{upenn1, upenn2}. \cite{selfish-shenker} and
\cite{under-over} took a game-theoretic approach to selfish route
control and to the relation between the overlay and the underlying
network, respectively. The theoretical frameworks proposed in
\cite{overlay-tcp, Kelly} formulated the problem of joint multi-path
route and rate control and provided a sufficient condition for the
stability of such decentralized algorithms.
\cite{chennee-cooexisting, dovrolis-infocom06} also demonstrate that
overlays can cause instability and \cite{dovrolis-infocom06} used
randomization to break synchronization.

In the media streaming community, the idea of path diversity is
traditionally combined with multiple-description coding:
complementary streams are simultaneously sent over independent
paths, to achieve resilience to loss in a bandwidth-efficient
manner. \cite{john-video-diversity} proposed to transmit multiple-
description video over independent paths; in follow-up work
\cite{john-cdn}, the same authors used this idea to design a
content-delivery network.
 \cite{yi-voice-diversity} applied the same idea to Voice-over-IP
 and also designed an playout scheduling algorithm to handle multi-path transmission.
 The same authors did a simulation study on the effect of
 replication and path diversity on TCP transfers \cite{steinbech-tcp-diversity}.

Our work fits in this scope as follows. It is related to
multi-homing and overlay approaches in that it tries to improve
end-to-end performance by connecting edge-networks via several
different ISPs and by exploiting their path diversity. We compare to
related work as follows. The novel aspect we are focusing on is
proactive replication of every packet over the available paths in a
single RAIL. This aspect is orthogonal to the online decision of
switching traffic between RAILs (i.e. sets of paths). However, in
this paper we still explore how to choose and manage the physical
paths that constitute a single RAIL. Similarly to \cite{upenn1,
upenn2}, we are looking at application-level metrics, particularly
for VoIP and TCP. In contrast to the media-streaming work, we
transmit redundant as opposed to complementary descriptions,
operating on the assumption that bandwidth is not the issue. Our
delay padding algorithm resembles playout buffering
\cite{yi-voice-diversity} in that it tries to smooth out the network
delay jitter; however, it is implemented at an edge device instead
of the end-point, and acts only as a playout-proxy without dropping
packets.

As the acronym ``RAIL'' indicates, our approach is inspired by the
{\em Redundant Array of Inexpensive Disks} (or {\em RAID}), an idea
for improving disk reliability and performance, proposed in the
classic SIGMOD'88 paper by G. Gibson and R. Katz \cite{raid}. The
basic idea of RAID was to combine multiple small, inexpensive disk
drives into an array of disk drives which yields performance
exceeding that of a Single Large Expensive Drive (SLED), and appears
to the computer as a single logical storage unit or drive.
Furthermore, disk arrays can be made fault-tolerant by redundantly
storing information in various ways. Five types of array
architectures, RAID-1 through RAID-5, were defined by the Berkeley
paper, each providing disk fault-tolerance and each offering
different trade-offs in features and performance. The different
levels of RAID in the original taxonomy \cite{raid} correspond to
various functions of an intelligent network device connected to
several ISPs. E.g. a network device that load-balances the outgoing
traffic over the available paths increases the throughput; it could
be named rail-0 because it corresponds to {\em striping}, or {\em
raid-level 0} in \cite{raid}. In this paper, we focus on packet
replication over several paths, which is analogous to {\em disk
mirroring}, or {\em raid-level 1} in \cite{raid}.

Similarly to RAID advocating multiple small inexpensive disks
instead of a single large expensive one, we believe that, as
bandwidth gets cheaper and cheaper, redundant replication of packets
over independent, inexpensive Internet connections becomes the
simplest, cost-efficient approach for achieving high
quality-of-service and reliability.

\section{System Design}
\label{sec:system}

\subsection{RAIL Mechanisms Overview}
\label{sec:rail-mechanisms}

RAIL improves the packet delivery between two remote local area
networks (LANs), by connecting them through multiple wide-area
paths. The paths are chosen to be as independent as possible, e.g.
belonging to different Internet Service Providers.
Fig.\ref{fig:topology} shows an example of two disjoint paths: Link
1 goes through ISP-A and ISP-C, Link 2 goes through ISP-B and ISP-D.
(The simplest configuration would be to have both LANs connected to
the same two ISPs.) For simplicity, we describe the system using two
paths only; the same ideas apply to $n>2$ paths.

A RAILedge device is required to connect each LAN to the wide-area
paths. Each packet that transitions from the LAN to the WAN, via the
RAILedge, is replicated at the RAILedge and sent out both WAN links.
Copies of the same packet travel in parallel through the different
WAN links and eventually arrive at the receiving RAILedge. There are
three possibilities: both copies arrive, one copy arrives or no copy
arrives.  The receiving RAILedge examines every packet coming in
from the WAN and suppresses any duplicates; i.e. it forwards the
first copy of each packet toward its destination but it discards any
copies arriving later.

The result is clear: the probability of both copies being lost is
reduced compared to using a single path, and the delay experienced
is the minimum of the delay on each path. Overall, the application
perceives a virtual RAIL link that is better than the underlying
physical links.

\begin{figure}[t]
\begin{center}
\centerline{\includegraphics[scale=0.25,angle=-90]{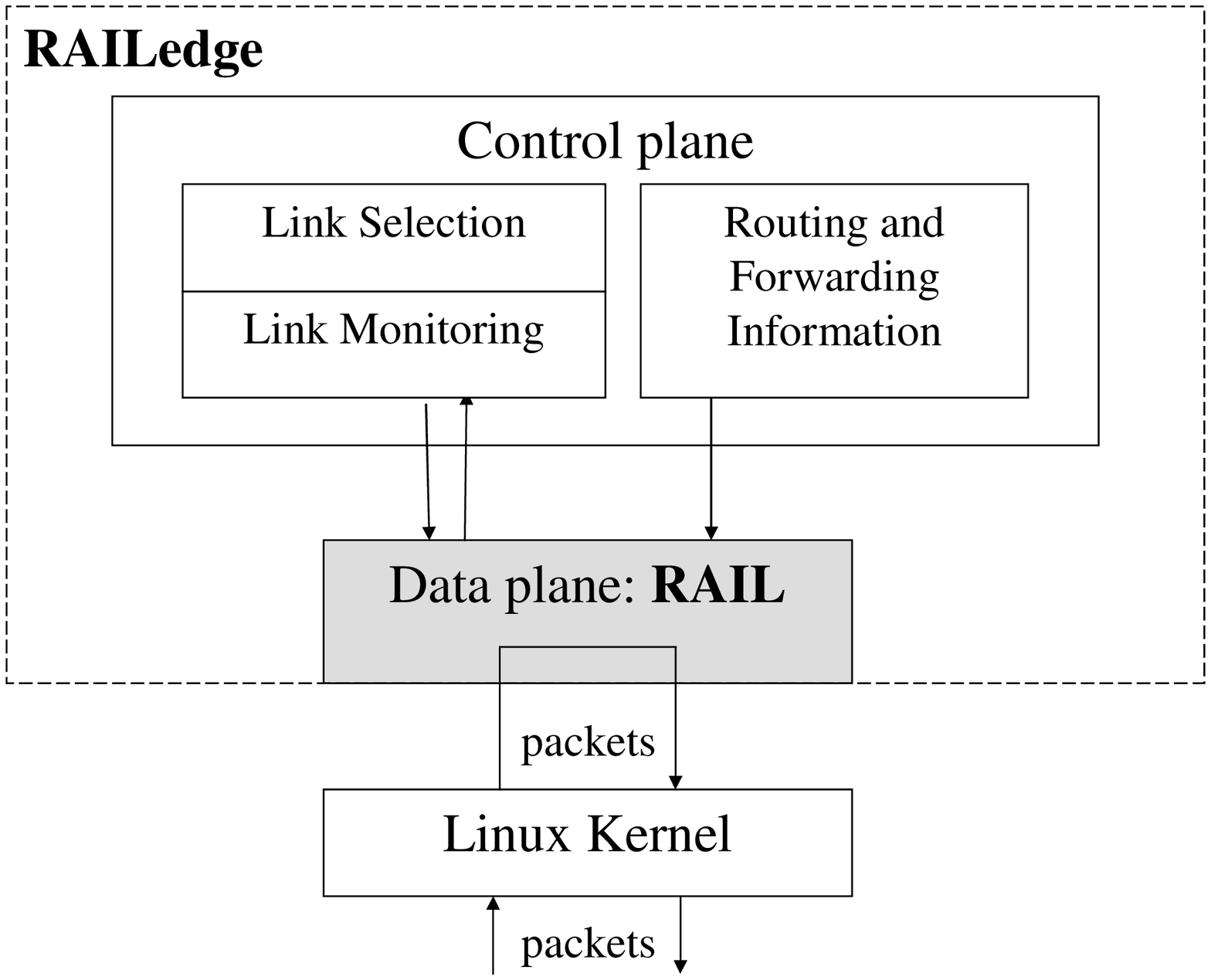}}
\end{center}
\vspace{-10pt} \caption{Components of our prototype RAILedge.}
\label{fig:railedge}
\end{figure}

\begin{figure}
{\centering \subfigure[Packet from LAN to WAN.]
{\includegraphics[scale=0.3,angle=-90]{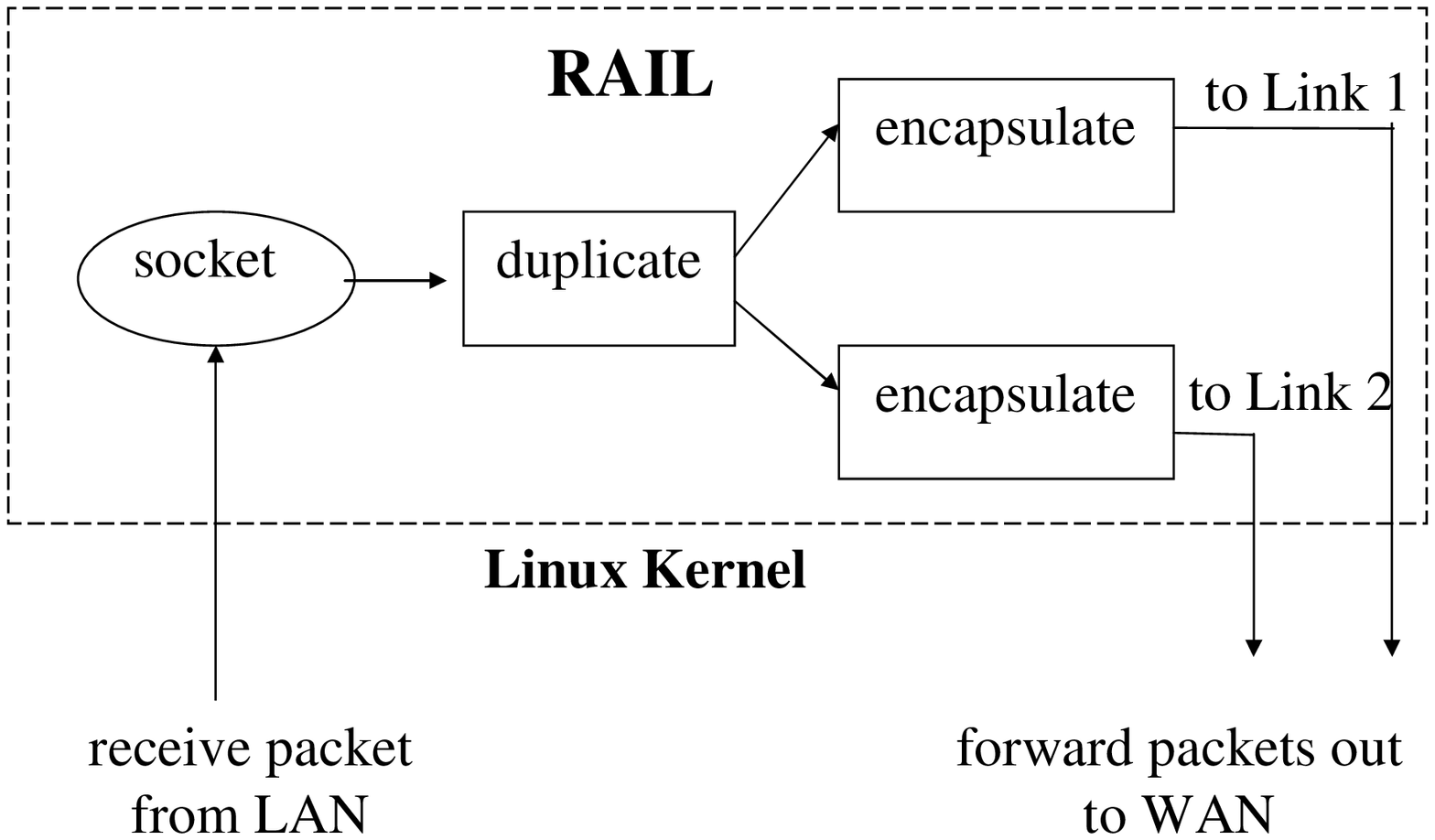}}
 \subfigure[Packet WAN to LAN.]
{\includegraphics[scale=0.3,angle=-90]{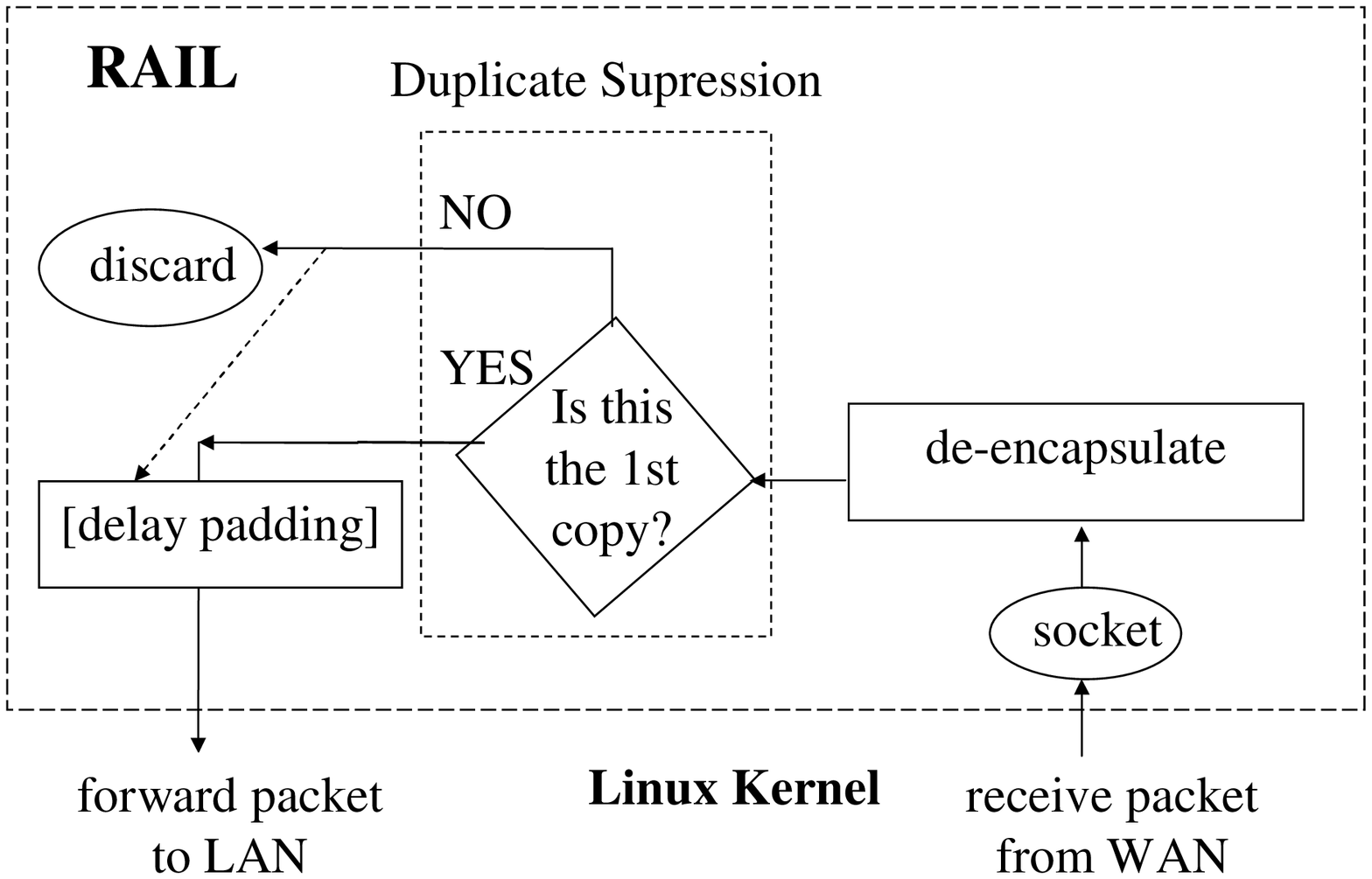}}
\caption{\label{fig:rail-lan-wan} RAIL functions upon reception of a
packet.}}
\end{figure}

In summary, the RAILedge performs three basic operations: (i) packet
duplication (ii) forwarding over all redundant Internet links and
(iii) duplicate suppression. RAILedge-RAILedge communication happens
over VPN tunnels, to ensure that every RAIL-ed packet is received by
the intended RAILedge. We implement tunneling with a simple
encapsulation/decapsulation scheme; our header includes the ID of
the sending RAILedge and a sequence number, which is used to
suppress duplicates at the receiving RAILEdge. All RAILedge
operations are transparent to the end-user. The components of a
RAILedge device are shown in Fig.\ref{fig:railedge} and the steps
taken upon reception of a packet are summarized in
Fig.\ref{fig:rail-lan-wan}.

There is a component of the RAILedge that we are not going to
examine in this paper: link monitoring and selection. This module is
responsible for monitoring the performance of every physical path,
computing appropriate quality metrics, and choosing the best subset
of paths to constitute the RAIL, over which packets should be
replicated. Link monitoring and dynamic selection is a research
problem in itself, with extensive and growing literature. In this
paper, we do not study dynamic path switching.\footnote{Intuitively,
we expect that dynamic RAIL switching is a less constrained problem
than single-path switching because (i) redundant transmission in a
single RAIL provides robustness to short-term problems and (ii) most
paths have consistent behavior in the longer time scales.} Instead,
we focus on (i) evaluating the replication of packets over {\em all}
paths that constitute the RAIL under study and (ii) on giving
recommendations on how to statically select these paths. This is
still useful for a typical use of RAIL: initially, the user compares
different ISPs and decides which is the best set to subscribe to;
after subscription, the user replicates packets over all ISPs.

\subsection{Delay Padding}
\label{sec:padding-mechanism}

Delay Padding is a mechanism that needs to complement the basic RAIL
mechanism  when there is delay disparity in the paths. The idea is
the following. The default behavior of the receiving RAILedge is to
forward the first copy and discard all copies that arrive later.
However, this may not always be the best choice when there is
significant delay disparity between the two paths. In such cases,
one can construct pathological scenarios where the default RAIL
policy results in patterns of delay jitter that adversely affect the
application.

One example is VoIP: the playout buffering algorithm at the receiver
tries to estimate the delay jitter and and adapt to it. This playout
algorithm is unknown to us and out of our control; even worse, it is
most likely designed to react to delays caused by real single paths,
not by virtual RAIL paths. For example, when path 1 is much faster
than path 2, then most of the time RAIL will forward copies arriving
from path 1. The playout buffer may adapt and closely match it, by
choosing a playout deadline slightly above the delay of the path 1.
When packets are lost on the fast path, the copies arriving from the
slow path will arrive late to be played out - and will be useless.
In this scenario, a better use of the two paths would be to
``equalize'' the delay in the two paths by artificially delaying the
packets arriving from the fast path, thus the name ``delay
padding''. Essentially, delay padding acts as a proxy for playout,
located at the RAILedge, and presents the receiver with the illusion
of a roughly constant one-way delay. The main differences from a
playout algorithm at the end-host is that delay padding does not
drop packets that arrive late for playout.

\begin{figure}
\begin{center}
\centerline{\includegraphics[scale=0.35]{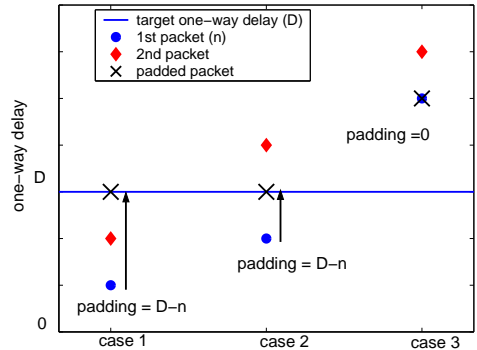}}
\end{center}
\vspace{-20pt} \caption{Delay Padding: artificially delay some
packets so that all packets experience the same one-way delay.}
\label{fig:paddingAlg}
\end{figure}

Fig. \ref{fig:paddingAlg} demonstrates the main idea of delay
padding, for packets in the same VoIP flow. The goal is to minimize
jitter, i.e. to make all packets experience the same, roughly
constant, one-way delay $D$, shown in straight line. For every
packet $i$, two copies arrive: the first one is marked with a
circle, the second is marked with the diamond. The actual time RAIL
forwards the packet is marked with an "X". Without padding, RAIL
would normally forward the first copy, which incurred one-way delay
$n_{RAIL}=min\{delay 1, delay2\}$. With padding, we compare
$n_{RAIL}$ to the target one-way delay $D$.
\begin{itemize}
\item In cases 1 and 2: $n_{RAIL}<D$. We wait for additional
``padding'' time $D-n_{RAIL}$ before forwarding the packet.
\item  In case 3: $n_{RAIL}>D$. We forward the packet immediately, without
further delay. (Instead, a playout algorithm at the receiver would
just drop the late packets).
\end{itemize}

The target one-way delay $D$ so as to maximize the overall voice
quality (MOS): $D=argmax\{MOS(D_{one_way})\}$. $D$ should be chosen
taking into account the statistics of two paths and the delay
budget. Adaptation of this value should be allowed only in much
larger time scales. We discuss the choice of $D$ to optimize $MOS$,
as well as the performance improvement from delay padding, in the
section on VoIP evaluation (\ref{sec:voip-quality}).

Delay padding may prove a useful mechanism for TCP as well. For
example, it could be used to remove reordering, caused by RAIL for
certain combinations of paths. This is discussed further in the
section on reordering (\ref{sec:reordering-general}) and in the
section on the effect of reordering on TCP in particular
(\ref{sec:reordering}).

A practical implementation of delay padding for VoIP would require
(i) the ability to identify voice packets and keep per-flow state
and (ii) calculations of timing in term of relative relative instead
absolute one-way delay. An implementation of reordering-removal for
TCP, would not necessarily require per flow state; it could just use
the sequence numbers on the aggregate flow between the two
RAILedges.

\subsection{RAIL Prototype and Experimental Setup}

In order to evaluate RAIL performance, we developed a RAILedge
prototype that implements the functionality described in Section
\ref{sec:rail-mechanisms}. Our prototype runs on Linux and consists
of a control-plane and a data-plane agent, both running in user
space. All routing and forwarding functionality is provided by the
Linux kernel. The control plane is responsible for configuring the
kernel with static routes and network interfaces. The data plane is
responsible for the packet processing, i.e.
encapsulation/decapsulation, duplication, duplicate suppression and
delay padding. In particular, the kernel forwards each received
packet to the data-plane agent, which processes it appropriately and
forwards it back to the kernel for regular IP forwarding, see
Fig.\ref{fig:railedge}.

Our user-space prototype is sufficient for a network connected to
the Internet through a T1 or T3 line: Without considering duplicate
packets, RAILedge running on a 1.9 GHz CPU with 512 MB of DRAM
forwards up to 100,000 minimum-size packets per second (about 51
Mbps) and up to 62,500 average-size (400 bytes) packets per second
(about 200 Mbps), while it introduces negligible jitter. For
higher-end links, we would need a different prototype that
implements the entire data path in kernel space.

\begin{figure}[t]
\begin{center}
\centerline{\includegraphics[scale=0.3,angle=-90]{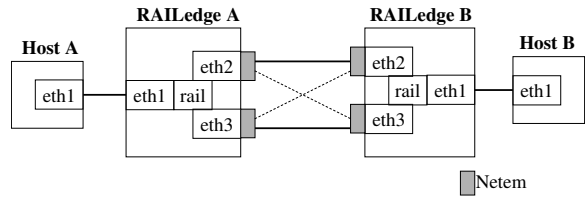}}
\end{center}
\vspace{-20pt} \caption{Experimental setup for RAIL}
\label{fig:testbed-RAIL}
\end{figure}

Fig. \ref{fig:testbed-RAIL} shows our experimental setup. Two Linux
boxes, Host-A and Host-B, communicate through prototype RAILedges A
and B respectively. The two RAILedges are connected directly through
two of their physical interfaces (eth2-eth2, eth3-eth3), thus
simulating the wide-area Links 1 and 2 shown in
Fig.\ref{fig:topology}.

We used Netem \cite{netem} on interfaces eth2, eth3, to
 emulate the properties of wide-area
networks in a controlled way. The current version of Netem emulates
variable delay, loss, duplication and re-ordering. Netem is
currently enabled in the Linux kernel. We also emulated WAN links of
various bandwidths, using the rate limiting functionality in Linux
(iproute2/tc).

\section{Performance evaluation}
\label{sec:evaluation}

In section \ref{sec:network-level}, we show that RAIL outperforms
any of the underlying physical paths in terms of network-level
metrics, i.e. it reduces loss, delay/jitter, it improves
availability and it does not make reordering any worse than it
already is in the underlying paths. In sections \ref{sec:voip} and
\ref{sec:tcp} we look at the improvement in terms of
application-level metrics for VoIP (MOS) and TCP (throughput); we
also look at how this improvement varies with the characteristics,
combinations and number of underlying paths.

\subsection{RAIL improves network-level metrics}
\label{sec:network-level}

RAIL statistically dominates any of the underlying paths, i.e. it
presents the end-systems with a virtual path with better statistics
in terms of network-level metrics (loss, delay, jitter and
availability). This is intuitively expected. At the very least, RAIL
could use just one of the paths and ignore the other; having more
options should only improve things. A natural consequence is that
any application performance metric calculated using these statistics
(e.g. loss rate, average delay, jitter percentiles) should also be
improved by RAIL; we found this to be indeed the case in computing
metrics for VoIP and TCP. In addition to the statistics, we also
looked at pathological sample paths, e.g. that reordering or special
patterns of jitter may arise; we show that RAIL does not make things
worse than they already are and that delay padding is able to handle
these cases.

\subsubsection{Loss}

Clearly, RAIL decreases the average packet loss rate from $p_1$,
$p_2$ to $p=p_1p_2$, for independent paths. One can derive some
useful rules of thumb, based on this simple fact.

{\em Number of paths.} Given that the actual loss rates are really
small $p_i<<0.1$ in practice, every new independent reduces loss
$p=p_1p_2...p_n$, by at least an order of magnitude. For similar
paths ($p_1=...p_n=p$) and it is easy to see that the loss
probability $P_{RAIL}(k)=p^k$ is a decreasing and convex function of
the number of paths ($k$). Therefore, most of the benefit comes from
adding the $2^{nd}$ path, and additional paths bring only decreasing
returns. However, adding a second path with significant {\em
different} (smaller) loss rate dominates the product and makes a big
difference.

\begin{figure}
\begin{center}
\centerline{\includegraphics[scale=0.45]{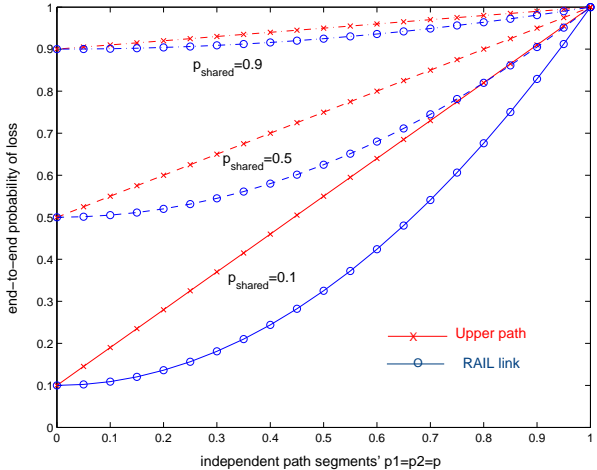}}
\end{center}
\vspace{-30pt} \caption{{\em The effect of shared loss.} Consider
two paths with shared loss rate $p_{shared}$ and independent loss
each $p_1=p_2=p$.  Here we plot the end-to-end $p_{RAIL}=$ and
$p_{single}$ vs. $p$, for various values of $p_{shared}$.}
\label{fig:shared-gain}
\end{figure}

{\em Correlation.} In practice, the physical paths underlying RAIL
may overlap. E.g. consider two paths that share a segment with loss
rate $p_{shared}$, and also have independent segments with
$p_1=p_2=p$. Loss experienced on any of the single paths w.p.
$p_{single}=(1-p)(1-p_{shared})$. Loss is experienced over RAIL w.p.
$p_{RAIL}=(1-p^2)(1-p_{shared})$. Fig. \ref{fig:shared-gain} plots
$p_{RAIL}$ vs. $p$ for various values of $p_{shared}$. Clearly,
$p_{RAIL}$ increases in both $p$ and $p_{shared}$. The lossier the
shared part, $p_{shared}$, compared to the independent part, $p$,
the less improvement we get by using RAIL (the curves for $p_{RAIL}$
and $p_{single}$ get closer and closer). Therefore, one should not
only look at their end-to-end behavior, but also at the quality of
their shared part, and choose a combination of paths that yields the
lowest overall $p_{RAIL}$.

RAIL also decreases the {\em burstiness in loss}. Due to lack of
space, we omit the analysis and refer the reader to section
\ref{sec:voip-testbed}, for testbed experiments that demonstrate
this fact.

\subsubsection{Availability}

The simplest way to view a ``failure''  is  as a long lasting period
of loss, and we can talk about the percentage of time a path spends
in failure. Then, the arguments we made for loss in the previous
section apply here as well. E.g. for RAIL to fail, both paths must
fail; the downtime reduces fast with the number and quality of
paths. Table \ref{table:Ken-downtime} gives a concrete idea on how
much RAIL decreases the downtime.
\begin{table}[h]
\begin{center}
\begin{tabular}{c|c}
If both Internet links have & Then the RAIL link has \\
that much {\em bad} time: & that much {\em medium} time: \\
\hline {\footnotesize 10\% (2.5 hours/day)} & {\footnotesize 1\% (1.5 hour/week)} \\
\hline {\footnotesize 2\% (3+ hours/week)} & {\footnotesize 0.04\% (2 hours/week)}\\
\hline {\footnotesize 0.5\% (1- hours/week)} & {\footnotesize 0.0025\% (15 min/year)} \\
\hline {\footnotesize 0.1\% (5 hours/month)} & {\footnotesize (2.5
0.0001\% (20 sec/year)}
\end{tabular}
\caption{\label{table:Ken-downtime}RAIL reduces downtime {\em and}
improves quality}
\end{center}
\end{table}

Note that RAIL not only reduces the time we spend in a ``bad
period'', but also improves the user experience from ``bad'' to
``medium'' during that period. We demonstrate this in detail in the
VoIP section (in particular see Table \ref{table:MOStable}).

\subsubsection{Delay and Jitter}
When a packet $i$ is RAIL-ed over two independent paths,  the two
copies experience one-way delay $d_1(i)$ and $d_2(i)$, and the
packet forwarded by RAIL (the copy that arrived first) experiences
$d(i)=min\{d_1(i), d_2(i)\}$. If the cumulative distribution
function (CDF) for $d_j,~j=1,2$ is $F_j(t)=Pr[d_i \le t]$, then the
delay CDF for RAIL is :
\begin{equation}
\begin{split}
F(t)=Pr[d\le t]=Pr[min\{d_1,d_2\}\le t]=...~~~~~~~\\
1-Pr[d_1> t~ and~d_2>t]=1-(1-F_1(t))(1-F_2(t))
\end{split}
\end{equation}
It is easy to see that RAIL statistically dominates any of the two
paths. Indeed, the percentage of packets experiencing delay more
than $t$ over RAIL is $1-F(t)=(1-F_1(t))(1-F_2(t))$, which is
smaller than the percentage of packets exceeding $t$ on any of the
two links ($1-F_i(t)$). This means that the entire delay CDF is
shifted higher and left, thus $F$ dominates $F_1$ and $F_2$. Any
quality metrics calculated based on these statistics (e.g. the
average delay, percentiles, etc) will be better for RAIL than for
any of the two paths. Rather than plotting arbitrary distributions
at this point, we choose to demonstrate the delay and jitter
improvement in some practical scenarios considered in the VoIP
section (\ref{sec:voip}).

\subsubsection{\label{sec:reordering-general}Reordering}

An interesting question is whether RAIL introduces reordering, which
may be harmful for TCP performance? In this section, we show that
RAIL does not make things worse than they already are on the
underlying paths. RAIL cannot reorder packets if each underlying
path does not reorder and does not drop packets. RAIL may translate
loss on one path to late arrivals from the other path, which is
actually an improvement.

\begin{figure}
\begin{center}
{\centering \subfigure[If each path does not reorder or drop
packets, then RAIL cannot reorder them.]
{\includegraphics[scale=0.3,angle=-90]{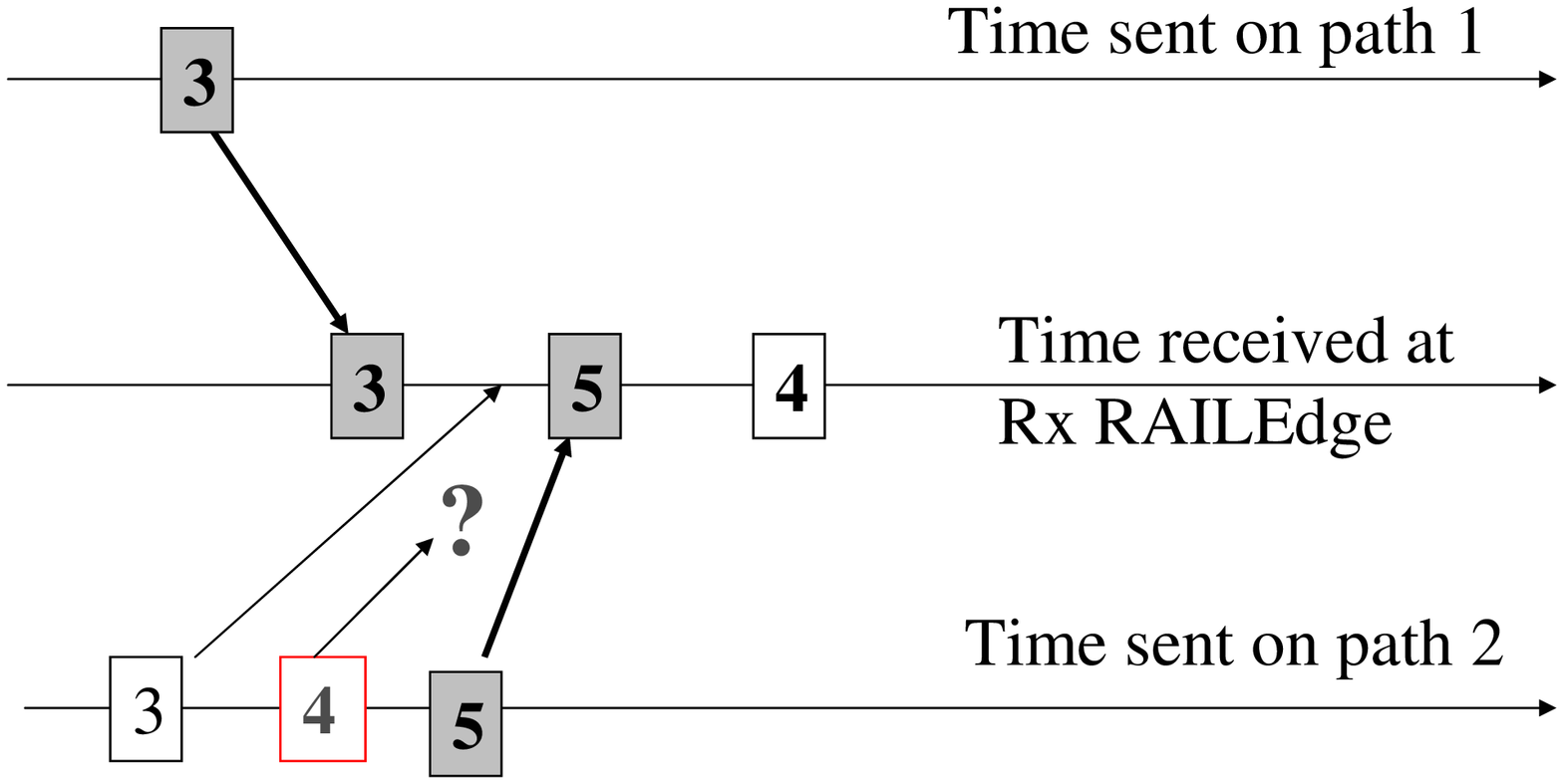}}}
{\centering \subfigure[RAIL converts loss on the faster path to
reordering, if a packet is dropped on the fast path and
$dt<d_2-d_1$.]
{\includegraphics[scale=0.3,angle=-90]{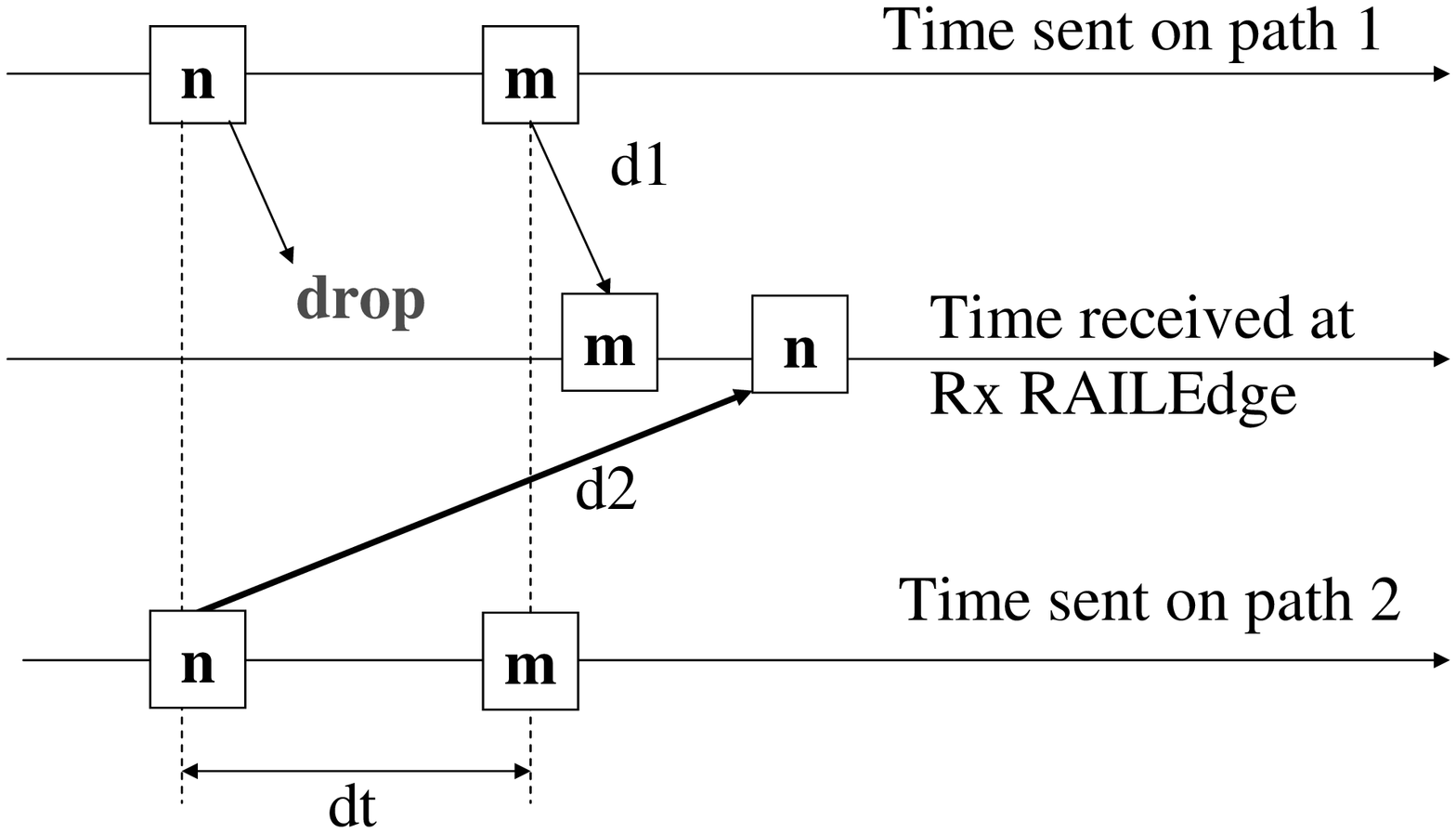}}}
\end{center}
\vspace{-10pt} \caption{\label{fig:reorder-example}RAIL and
Reordering}
\end{figure}

{\em {\bf Proposition 1.} If each path does not drop or
reorder packets, then RAIL cannot introduce reordering.}\\
{\bf Proof.} Let us assume  that RAIL can reorder packets.
Fig.\ref{fig:reorder-example}(a) shows an example out-of-order
sequence of out of order packets forwarded by the receiving
RAILedge: (3,5,4). The same arguments will hold for any sequence
$(i,k,j)$ with $i<j<k$. Packets 3 and 5 must have arrived through
different paths (otherwise one of the paths would have dropped
packet 4 or reorder it). Say 3 arrives from the top path and 5 from
the bottom path. Then the copy of 3 sent on the bottom path must
have arrived between 3 and 5 (otherwise RAIL would have forwarded
the bottom 3 copy first). What happened to packet 4 sent on the
bottom path? If it arrived between 3 and 5, then there would be no
out-of-order at RAIL; if it arrived after 5, then the bottom path
would have reordered 4 and 5, which we assumed it is not the case;
and we have assumed that 4 is not dropped either. We reached a
contradiction, which means that RAIL cannot reorder packets if both
paths are well behaving to start with.

{\em {\bf  Proposition 2.} RAIL may translate loss on the faster
path to late arrivals from the slower path. If the inter-packet
spacing at the sender is smaller than
the delay difference of the two paths, then the packets arrive out of order.} \\
{\bf Example.} In Fig.\ref{fig:reorder-example}(b), we consider
paths 1 and 2, with one-way delay $d_1<d_2$. Two packets $n$ and $m$
are sent with spacing $dt$ between them. If packet $n$ is lost on
the fast path, and $dt \le d_2-d_1$, then $n$ will arrive at the
RAILedge after $m$ and the RAILedge will forward them out-of-order.
The larger the delay difference $d_2-d_1$ and the smaller the
spacing between packets $dt$, the larger the reordering gap.

{\em {\bf Fact 3.} Better late than never.}\\
{\bf Discussion.} For VoIP, it does not hurt to receive packets
late, as opposed to not receive them at all.
However, out-of-order packets may potentially hurt TCP performance.
Testbed experiments, in section \ref{sec:reordering}, show that TCP
performs better when $x\%$ of packets out-of-order, compared to when
$x\%$ of packets lost. Furthermore, the delay padding component is
designed to handle the timely delivery of packets. We will revisit
this fact in section \ref{sec:reordering}.

\subsection{\label{sec:voip}RAIL improves VoIP performance}

\subsubsection{Voice-over-IP Quality}
\label{sec:voip-quality}

 A subjective measure used to assess Voice-over-IP
quality is the Mean Opinion Score (or MOS), which is a rating in a
scale from 1 (worst) to 5 (best) \cite{P.800}. Another equivalent
metric is the $I$ rating, defined in the Emodel \cite{g.107}.
\cite{g.107} also provides a translation between $I$ and $MOS$; in
this paper, we convert and present voice quality in the MOS scale
only, even when we do some calculations in the $I$ scale .

\begin{figure}
\begin{center}
\centerline{\includegraphics[scale=0.3]{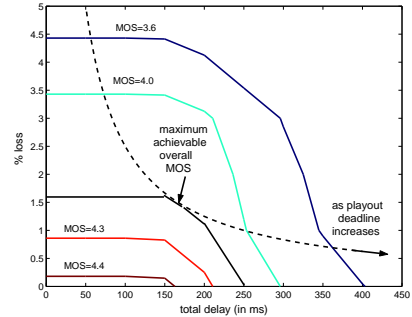}}
\end{center}
\vspace{-20pt} \caption{Voice Quality as a function of both loss and
delay.} \label{fig:MOScontours}
\end{figure}

VoIP quality has two aspects. The first is speech quality and it
depends primarily on how many and which packets are dropped in the
network and/or at the playout buffer. \cite{g.107,g.113} express the
speech quality as a function of the packet loss rate,
$MOS_{speech}(loss~rate)$, for various codecs. The second aspect of
VoIP quality is interactivity, i.e. the ability to comfortably carry
on an interactive conversation; \cite{g.107. g.114} express this
aspect as a function of the average one-way delay,
$MOS_{interactivity}(avg~delay)$, for various conversation types.
These two aspects can be added together (in the appropriate $I$
scale defined in \cite{g.107}) to give an overall MOS rating:
$MOS=MOS_{speech}+MOS_{interactivity}$. This is the metric we will
use throughout this section.

We do not present the details of these formulas in this submission,
due to lack of space. The interested reader is referred to the ITU-T
standards \cite{g.107, g.113, g.114} or to comprehensive tutorials
on the subject \cite{athina-voip,cole}. What the reader needs to
keep in mind is that there are either formulas or tables for
$MOS_{speech}(loss~rate)$, $MOS_{interactivity}(avg~delay)$ and that
$MOS=MOS_{speech}+MOS_{interactivity}$. This is a commonly used
methodology for assessing VoIP quality, e.g. see
\cite{athina-voip,upenn1}. Fig.\ref{fig:MOScontours} shows contours
of MOS as a function of loss and delay based on the data provided in
the ITU-T standards, considering G.711 codec and free conversation.

{\bf The effect of playout.} In the assessment of VoIP, one should
take into account the function of the playout algorithm at the
receiver, which determines the  playout deadline $D_{playout}$:
packets with one-way delay exceeding $D_{playout}$ are dropped. As
$D_{playout}$ increases, the one-way delay increases (thus making
interactivity worse), but less packets are dropped due to late
arrival for playout (thus making speech quality better). Therefore,
there is a tradeoff in choosing $D_{playout}$ and one should choose
$D_{opt}=argmax{MOS(D_{playout})}$. This tradeoff depicted in Fig.
\ref{fig:MOScontours} and is also responsible tfor the shape of the
$MOS(D_{one~way})$ curves of Fig.\ref{fig:mos-vs-delay}, which
clearly have a maximum at $D_{opt}$. The value $D_{opt}$ depends on
the loss, delay and jitter of the underlying paths as welllas on the
delay budget consumed in components other than the playout. Recall
that $D_{playout}$ is only a part of the total
$D_{one~way}=D_{end~systems}+D_{network}+D_{playout}$ and that
packets arriving late contribute to the total loss
($packet~loss=(network~loss)+Pr[d>D_{playout}]$).

{\em The effect of RAIL.} In the previous section, we saw that RAIL
decreases (i) the loss rate (ii) the average delay and (iii) the
percentage of late packets. Therefore, it also improves the $MOS$
which is a function of these three statistics.

\subsubsection{Railing VoIP over representative Internet Paths}

In this section, we now use realistic packet traces to simulate the
behavior of WAN links. In particular, we use the packet traces
provided in \cite{athina-traces}, which are collected over the
backbone networks of major ISPs, by sending probes that emulate
G.711 traffic. 

\begin{figure}
\begin{center}
\centerline{\includegraphics[scale=0.3]{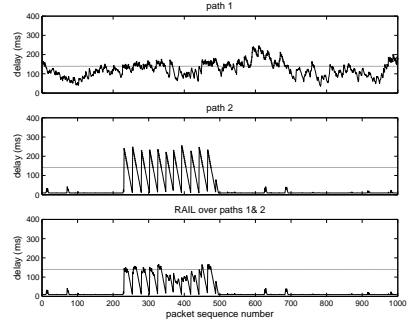}}
\end{center}
\vspace{-30pt} \caption{One-way delay experienced when packets are
transmitted over example path 1, example path 2, and using RAIL over
these two paths ($d_RAIL=min\{d_1,d_2\}$).} \label{fig:trace-b1-m2}
\end{figure}

Fig.~\ref{fig:trace-b1-m2}(a) and (b) show the delay experienced on
two paths between San Jose, CA and Ashburn, VA. The two paths belong
to two different ISPs and experience different delay patterns.
Fig.\ref{fig:trace-b1-m2}(c) shows the one-way delay experienced by
packets RAIL-ed over these two paths. Packets were sent every 10ms.

\begin{figure}
\begin{center}
\centerline{\includegraphics[scale=0.3]{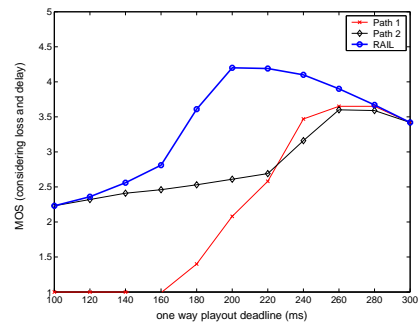}}
\end{center}
\vspace{-20pt} \caption{MOS vs. playout deadline for the traces in
Fig.\ref{fig:trace-b1-m2}.} \label{fig:mos-vs-delay}
\end{figure}

Although there is no network loss in these example paths, packets
may still be dropped if they arrive after their playout deadline.
Because the action of playout is out of the control of RAILedge, we
consider the entire range of fixed one-way playout deadlines (out of
which 70ms are considered consumed at the end-systems). The
resulting $MOS$ is shown in Fig.\ref{fig:mos-vs-delay} as a function
of $D_{one~way}$.\footnote{The curve $MOS(D_{one~way})$ has a
maximum which corresponds to $D_{playout}^{opt}$ that optimizes the
loss-delay tradeoff in the overall $MOS$.} Notice that the $MOS$
curve for RAIL is higher then both curves corresponding to
individual links, for the entire range of delays considered.

In general, RAIL always improves VoIP quality because it presents
the application with a better virtual path in terms of loss, delay
and jitter. However, the relative improvement of RAIL vs. the single
path depends (i) on the behavior of the two paths and (ii) on the
playout algorithm.

\begin{figure}[t!]
\begin{center}
\centerline{\includegraphics[scale=0.4]{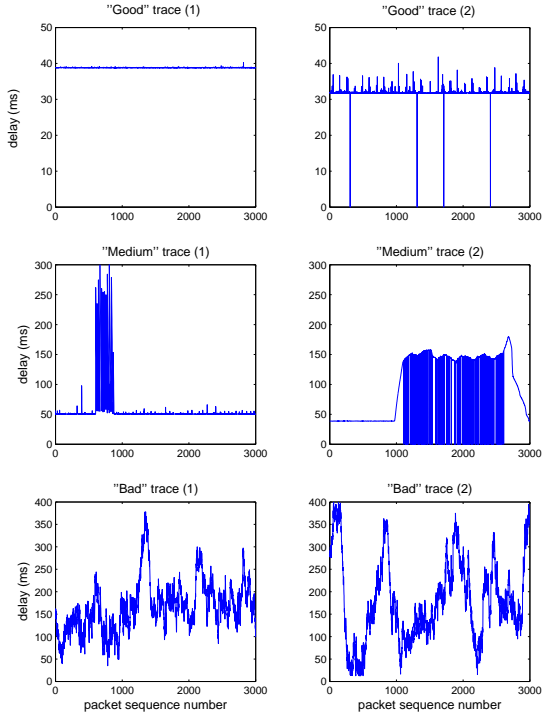}}
\end{center}
\vspace{-20pt} \caption{{\footnotesize Six representative packet
traces, collected over wide-area paths of Internet backbones
\cite{athina-traces}. We plot one-way delay vs. packet sequence
number; when a packet is lost we give it a 0 value.}}
\label{fig:alltraces} \vspace{-10pt}
\end{figure}

This was just an illustrative example of RAIL over two specific
paths. We now consider additional representative traces and their
combinations using RAIL. We consider six packet traces from
\cite{athina-traces}, shown in Fig.~\ref{fig:alltraces}. We call the
traces ``good'', ``medium'' and ``bad'', to roughly describe the
VoIP performance they yield.\footnote{E.g. we call the two traces on
the top row ``good'', because they have almost constant delay, and
negligible or no loss. We call the two traces on the medium row
``medium'' because they are good most of the time, except for a
period of high delay/jitter/loss. Finally, the traces in the bottom
row have very high delay (up to 400ms!) and jitter.}

We then considered pairs of paths for all the combinations of
good/medium/bad quality, by choosing one trace from the left and the
second trace from the right of Fig.\ref{fig:alltraces}. Table
\ref{table:MOStable} shows the MOS for each one of the 6 paths, as
well as for these 9 combinations using RAIL.\footnote{In all cases,
a conservative fixed playout deadline of 200ms was considered; 40ms
delay has also been added for the end-systems).} One can see that
the combined link (RAIL) provides one ``class'' better quality than
any of the individual links. For example, if at least one path is
good ($MOS>4$), then it dominates and the RAIL link is good,
regardless of the second link. Two medium links (roughly $3<MOS<4$)
give a good RAIL link($MOS>4$) and two bad links ($MOS<2$) give a
medium RAIL link, i.e. there is one class of service improvement.
This is intuitively expected, because RAIL multiplexes and uses the
best of both paths. In addition, we did in-house informal listening
tests: we simulated the transmission of actual speech samples over
these traces and we had people listen to the reconstructed sound. It
was clear that the RAIL-sample sounded much better.

\begin{table}[th!]
\begin{center}
\begin{tabular}{c||c|c|c}
{\bf RAIL}& \multicolumn{3}{c}{\bf Path 2}\\
\hline \multicolumn{1}{c||}{\bf Path 1} & Good-2 & Medium-2 & Bad-2 \\
\multicolumn{1}{c||}~ &(4.19) & (3.02) & (1.19) \\
\hline
\hline Good-1 & ~& ~& ~~\\
(4.21) & 4.21 &  4.21 & 4.21\\
\hline Medium-1 & ~& ~ & ~\\
(3.87) & 4.21 & 4.21 & 4.00\\
\hline Bad-1 & ~& ~& ~~\\
(1.76) & 4.20& 3.97 & 3.09\\
\end{tabular}
\caption{\label{table:MOStable}Voice Quality (in terms of MOS score)
for the 6 representative paths, and for their 9 combinations using
RAIL.}
\end{center}
\vspace{-10pt}
\end{table}

Notice, that this quality improvement is in addition to the
availability improvement in Table \ref{table:Ken-downtime}: not only
RAIL reduces the time spent in ``bad/medium'' periods, but it also
improves the experience of the user during that period, from ``bad''
to ``medium'' and from ``medium'' to ``good''.

\subsubsection{\label{sec:voip-testbed}Testbed experiments for VoIP-over-RAIL}

In this section, we use our testbed to demonstrate the improvement
that RAIL brings to VoIP quality for the entire range of path
conditions. We used Netem to control the loss and delay parameters
of each path. We sent probes to emulate the transmission of voice
traffic.\footnote{200B UDP packets were sent every 20ms
(corresponding to G.711 at 64kbps and 20ms packetization: 160B
payload and 40B RTP/UDP/IP headers) for 2min duration.}

\begin{figure}
{\centering \subfigure[Packet loss measured on each path.]
{\includegraphics[scale=0.3]{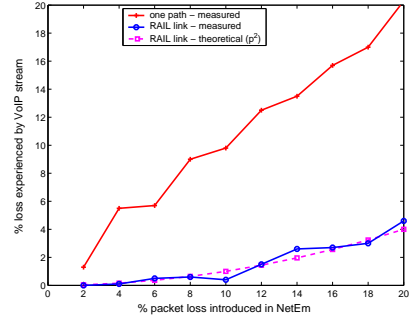}}
 \subfigure[Resulting MOS (speech quality)]
{\includegraphics[scale=0.3]{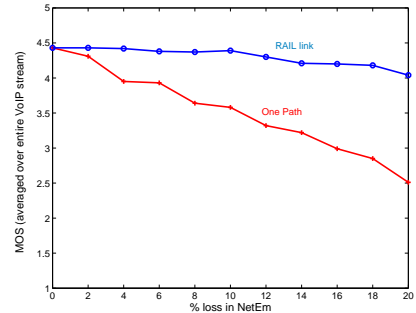}}
\caption{\label{fig:loss-testbed} Testbed experiments on the effect
of packet loss to VoIP with/without RAIL. }}
\end{figure}

First, we looked at {\em loss rate}. We applied uniform loss and the
same loss rate $p$ from  1 to 20\%, which is quite high but may
happen during short periods of bursty loss. As expected, the voice
stream experiences loss rate $p^2$ when transmitted over RAIL, and
$p$ over on a single link. Indeed, in Fig.\ref{fig:loss-testbed}(a),
the measured $45$ degrees red line (for a single link) agrees with
$p$; the measured blue line (for RAIL) agrees with the theoretical
$p^2$ dashed purple line. This loss reduction results in a speech
quality improvement up to 1.5 units of MOS. Fig.
\ref{fig:loss-testbed}(b) shows that MOS (averaged over the entire
duration) is practically constant when we use RAIL, while the MOS
over a single link is decreasing rapidly with increasing loss rate.
A side-benefit is that speech quality varies less with time, which
is less annoying for the user.


\begin{table}[h!]
\begin{center}
\begin{tabular}{c||c|c|c|c|c}
\hline \multicolumn{6}{c} {\bf Number of packets lost in burst} \\
\hline
{\bf {\footnotesize Loss}}& \multicolumn{5}{c}{\bf Loss Rate}\\
{\bf {\footnotesize Corr.} }& 10\% & 20\% & 30\% &40\% & 50\% \\ 
\hline\hline
{\footnotesize 0\%}  &   {\footnotesize 99/{\bf 11}}&  {\footnotesize 203/{\bf 58}} & {\footnotesize 298/{\bf 101}} & {\footnotesize 399/{\bf 160}} &{\footnotesize 514/{\bf 249}}  \\
{\footnotesize 20} &   {\footnotesize 27/{\bf 1}} &  {\footnotesize 127/{\bf 14}} & {\footnotesize 257/{\bf 62}}  & {\footnotesize 362/{\bf 158}} &{\footnotesize 512/{\bf 242}}  \\
{\footnotesize 40} &   {\footnotesize 6/{\bf 0}}  &  {\footnotesize 45/{\bf 1}}   & {\footnotesize 144/{\bf 33}}  & {\footnotesize 340/{\bf 129}} &{\footnotesize 479/{\bf 251}}  \\
{\footnotesize 60} &   {\footnotesize 0/{\bf 0}}  &  {\footnotesize 18/{\bf 0}}   & {\footnotesize 76/{\bf 8}}    & {\footnotesize 248/{\bf 82}}  &{\footnotesize 537/{\bf 258}}  \\
{\footnotesize 80} &   {\footnotesize 0/{\bf 0}}  &  {\footnotesize 0/{\bf 0}}    & {\footnotesize 14/{\bf 0}}    & {\footnotesize 123/{\bf 12}}  &{\footnotesize 466/{\bf 288}}  \\
\hline
\end{tabular}
\caption{\label{table:inburst}Number of packets lost in burst (out
of 1000 total) on a single path (shown in regular font) vs. RAIL
(shown in bold font).}
\end{center}
\vspace{-15pt}
\end{table}
\begin{table}[h!]
\begin{center}
\begin{tabular}{c||c|c|c|c|c}
\hline \multicolumn{6}{c} {\bf Number of bursts} \\
\hline
{\bf {\footnotesize Loss}}& \multicolumn{5}{c}{\bf Loss Rate}\\
{\bf {\footnotesize Corr.} }& 10\% & 20\% & 30\% &40\% & 50\% \\
\hline\hline
{\footnotesize 0\%} & {\footnotesize 88/{\bf 11}}&  {\footnotesize 161/{\bf 52}} & {\footnotesize 204/{\bf 93}} & {\footnotesize 243/{\bf 137}} &{\footnotesize 237/{\bf 197}}   \\
{\footnotesize 20\%}& {\footnotesize 22/{\bf 1}}&   {\footnotesize 93 /{\bf 13}} & {\footnotesize 185/{\bf 52}} & {\footnotesize 197/{\bf 122}} &{\footnotesize 230/{\bf 178}}  \\
{\footnotesize 40\%}& {\footnotesize 5/{\bf 0}}&    {\footnotesize 37/{\bf 1}}   & {\footnotesize 99/{\bf 28}}  & {\footnotesize 175/{\bf 90}}  &{\footnotesize 198/{\bf 159}}   \\
{\footnotesize 60\%}& {\footnotesize 0/{\bf 0}}&    {\footnotesize 13/{\bf 0}}   & {\footnotesize 50/{\bf 7}}   & {\footnotesize 124/{\bf 57}}  &{\footnotesize 164/{\bf 145}}  \\
{\footnotesize 80\%}& {\footnotesize 0/{\bf 0}}&    {\footnotesize 0/{\bf 0}}    & {\footnotesize 4/{\bf 0}}    & {\footnotesize 53/{\bf 7}}    &{\footnotesize 100/{\bf 97}}   \\
\hline
\end{tabular}
\caption{\label{table:numburst}Number of bursts (out of 1000) on a
single path (in regular font) vs. RAIL (in bold).}
\end{center}
\vspace{-15pt}
\end{table}
\begin{table}[h!]
\begin{center}
\begin{tabular}{c||c|c|c|c|c}
\hline \multicolumn{6}{c} {\bf Maximum burst size} \\
\hline
{\bf {\footnotesize Loss}}& \multicolumn{5}{c}{\bf Loss Rate}\\
{\bf {\footnotesize Corr.} }& 10\% & 20\% & 30\% &40\% & 50\% \\
\hline\hline
{\footnotesize 0\%}  &   {\footnotesize 2/{\bf 1}}&  {\footnotesize 5/{\bf 4}} & {\footnotesize 7/{\bf 2}} & {\footnotesize 7/{\bf 3}} &{\footnotesize 11/{\bf 3}}  \\
{\footnotesize 20\%}  &   {\footnotesize 3/{\bf 1}}&  {\footnotesize 4/{\bf 2}} & {\footnotesize 5/{\bf 3}} & {\footnotesize 8/{\bf 7}} &{\footnotesize 17/{\bf 5}}  \\
{\footnotesize 40\%}  &   {\footnotesize 2/{\bf 0}}&  {\footnotesize 3/{\bf 1}} & {\footnotesize 8/{\bf 4}} & {\footnotesize 6/{\bf 5}} &{\footnotesize 10/{\bf 7}}  \\
{\footnotesize 60\%}  &   {\footnotesize 0/{\bf 0}}&  {\footnotesize 2/{\bf 0}} & {\footnotesize 4/{\bf 2}} & {\footnotesize 10/{\bf 4}} &{\footnotesize 19/{\bf 7}} \\
{\footnotesize 80\%}  &   {\footnotesize 0/{\bf 0}}&  {\footnotesize 0/{\bf 0}} & {\footnotesize 10/{\bf 0}} & {\footnotesize 8/{\bf 2}} &{\footnotesize 24/{\bf 16}} \\
\hline
\end{tabular}
 \caption{\label{table:maxburst}Maximum size of burst
(i.e. max number of consecutive packets lost) on a single path (in
regular font) vs. RAIL (in bold font). The average burst size for
RAIL is 1 in most cases.}
\end{center}
\vspace{-15pt}
\end{table}

Second, we looked at the {\em burstiness of loss}, which is an
important aspect because it can lead to loss of entire phonemes,
thus degrading speech intelligibility. 
To control burstiness, we controlled the ``correlation'' parameter
in Netem.\footnote{The Netem correlation coefficient does increase
the loss burstiness, but does not directly translate to burstiness
parameters, such as burstiness duration. An artifact of their
implementation \cite{netem} is that increasing correlation decreases
the measured loss rate (for loss rate<50\%). However, it does not
matter: our point is to compare RAIL to a single path, under the
same loss conditions} We tried all combinations of
$(loss~rate,~loss~correlation)$ and measured the following metrics
for bursty loss: (i) number of packets lost in burst (ii) number of
bursts (iii) average burst size (iv) maximum burst size. In Tables
\ref{table:inburst},\ref{table:numburst}, \ref{table:maxburst}, we
show the numbers measured over one link in regular font, and the
numbers measured over RAIL in bold. Clearly, all metrics are
significantly reduced with RAIL compared to the single path case,
which demonstrates that RAIL reduces loss burstiness. This good
property is intuitively expected, as it is less likely that both
paths will experience a burst at the same time.

Third, we experimented with {\em delay jitter}. We considered two
paths with the same mean delay (100ms), and we used Netem to
generate delay according to a paretonormal distribution. We
generated delay on both paths according to the same statistics. We
fixed the mean delay at 100ms for both paths, and experimented with
the entire range of delay variability (standard deviation from 10ms
to 100ms and delay correlation from 0\% to 100\%).

\begin{figure}
\begin{center}
\centerline{\includegraphics[scale=0.35]{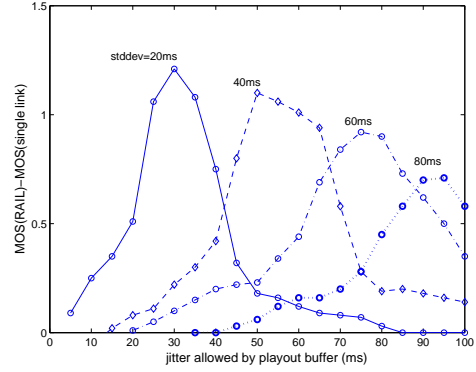}}
\end{center}
\vspace{-30pt} \caption{{\footnotesize Improvement in speech quality
using RAIL vs. using a single path, considering the full range of
two factors: (i) the delay variability of the underlying paths
(captured here by the standard deviation  of delay) and (ii) the
playout at the receiver (captured here by the jitter allowed). Delay
was configured in Netem to be paretonormal distributed, with
mean=100ms and correlation=0.}} \label{fig:jitter-improvement}
\end{figure}

In the beginning, we set delay correlation at 0 and increase the
standard deviation of delay. We observed that RAIL reduces the
jitter experienced by the VoIP stream. This results in less packets
being late for playout and thus better speech quality. The exact
improvement depends (i) on the delay variability of the underlying
paths (captured here by the standard deviation of delay) and (ii) on
the playout at the receiver (captured here by the jitter allowed at
the playout).

Fig.\ref{fig:jitter-improvement} shows the improvement in speech
quality (in MOS) compared to a single path, for a range of these two
parameters ($std~ dev$ 20-80ms and jitter level acceptable at
playout 20-100ms). One can make several observations. First, RAIL
always help (i.e. benefit$>0$); this is because RAIL presents the
end-system with a better virtual path. Second, there is a maximum in
every curve (every curve corresponds to a certain path delay
variability): when the playout is intolerant to jitter, then it
drops most packets anyway; when the playout can absorb most of the
jitter itself, then the help of RAIL is not needed; therefore, RAIL
provides most of its benefit, in the middle - when it is needed to
reduce the perceived jitter below the acceptable threshold for
playout. Finally, the entire curve  moves to the right and lower for
paths with higher delay variability.

In addition, we experimented with delay correlation (which will
result in several consecutive packets arrive late and get dropped in
the playout) and we observed that RAIL decreased this correlation by
multiplexing the two streams. Finally, we experimented with RAIL-ed
VoIP and several non-RAILed TCP flows interfering with it. The idea
was to have loss and delay caused by cross-traffic rather than being
artificially injected by Netem. RAIL brought improvement in the same
orders of magnitude as observed before.

\subsubsection{Delay Padding}

\begin{figure}[t!]
\begin{center}
\centerline{\includegraphics[scale=0.35]{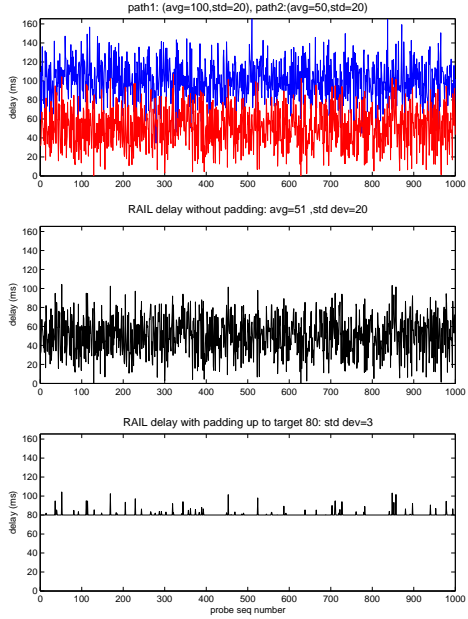}}
\end{center}
\vspace{-20pt} \caption{Padding decreases jitter for RAIL over two
paths with different average delay ($100ms$ and $50ms$) and similar
delay variability (e.g. $std dev=20ms$ for both).}
\label{fig:padding-diffavg}
\end{figure}

\begin{figure}[h]
\begin{center}
\centerline{\includegraphics[scale=0.35]{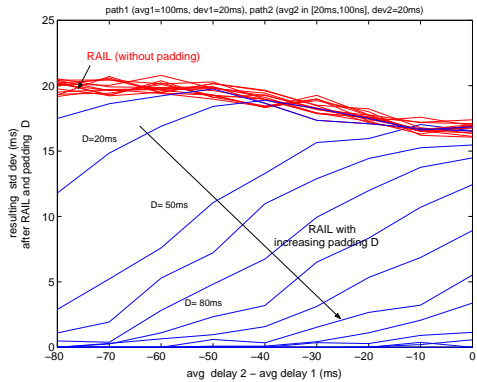}}
\end{center}
\vspace{-20pt} \caption{The larger the delay disparity between the
two paths, the more padding is needed.
} \label{fig:padding-delta-avg}
\end{figure}

\begin{figure}[t!]
\begin{center}
\includegraphics[scale=0.35]{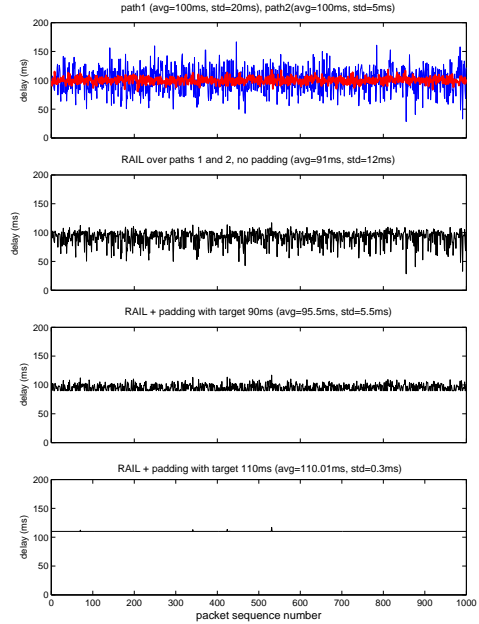}
\end{center}
\vspace{-20pt} \caption{Padding decreases jitter for RAIL over paths
with the same average delay ($100ms$) but different jitter
($stddev=20ms,5ms$). The more padding - the less jitter.}
\label{fig:padding-sameavg}
\end{figure}

The delay padding algorithm, described in section
\ref{sec:padding-mechanism}, acts as a proxy playout at the
receiving RAILedge: it artificially adds delay (``padding'') in
order to create the illusion of constant one-way delay.
In this section, we use matlab simulation to demonstrate the effect
of padding. Fig.\ref{fig:padding-diffavg} considers the case when
the two paths differ in their average delay; this can be due to e.g.
difference in propagation and/or transmission delay. Notice the
difference between (b)-RAIL without padding and (c)-RAIL with
padding. Fig.\ref{fig:padding-delta-avg} shows that the larger the
disparity between the two paths, the more padding is needed to
smooth out the stream. Fig. \ref{fig:padding-sameavg} considers the
case when two paths have the same average delay but differ
significantly in the delay jitter, e.g. due to different
utilization. Fig. \ref{fig:padding-sameavg}(a) plots the delay on
the two paths on the same graph; Fig. \ref{fig:padding-sameavg}(b)
shows what RAIL does without padding; Fig.
\ref{fig:padding-sameavg}(c) and (d) show that the stream can be
smoothed out by adding more padding. The appropriate amount of
padding should be chosen so as to maximize the overall MOS - as
discussed in section \ref{sec:voip-quality}.

\subsection{\label{sec:tcp}RAIL improves TCP performance}

In the section \ref{sec:network-level}, we saw that RAIL
statistically dominates the underlying paths in terms network-level
statistics. Therefore, performance metrics computed based on these
statistics, such as the average throughput, should be improved. In
section \ref{sec:analysis-tcp}, we analyze the throughput of
long-lived TCP flows, and we show that indeed this is the case.
However, there may be pathological cases, e.g. when reordering
falsely triggers fast-retransmit; this is what we study in section
\ref{sec:reordering}, and show that -for most practical cases- RAIL
helps TCP as well .

\subsubsection{\label{sec:analysis-tcp}Analysis of long-lived TCP-over-RAIL}

{\em A simple formula.} Let us consider two paths with loss rate and
round-trip times: ($p_1$, $RTT_1$), $(p_2, RTT_2)$ respectively, and
w.l.o.g. $RTT_1 \le RTT_2$. 
The simple rule of thumb from \cite{floyd} predicts that the
long-term TCP throughput for each path is:
$T_i=\frac{1.22}{RTT_i\sqrt{p_i}}, \text{for }i=1,2$. What is the
long-term TCP throughput using RAIL over these two paths? Following
a reasoning similar to \cite{floyd}, we find that: \vspace{-10pt}
\begin{equation}
T=\frac{1.22}{E[RTT]\sqrt{p_1p_2}}, \text{where:}~~~~~~~~~~~~\\
\label{eq:tcp-rail-formula}
\end{equation}
\begin{equation}
E[RTT]=RTT_1\frac{1-p_1}{1-p_1p_2}+RTT_2\frac{p_1(1-p_2)}{1-p_1p_2}
\label{eq:tcp-E[RTT]}
\end{equation}

\begin{figure}
\begin{center}
\centerline{\includegraphics[scale=0.35,angle=-90]{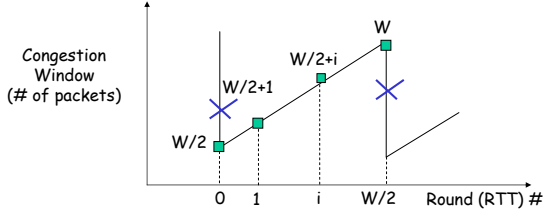}}
\end{center}
\vspace{-20pt} \caption{The simple steady-state model for TCP
\cite{floyd}.} \label{fig:tcp-prioni} \vspace{-10pt}
\end{figure}

{\bf Proof.} Fig. \ref{fig:tcp-prioni} shows the simple steady-state
model considered in \cite{floyd}. The network drops a packet from
when the congestion window increases to $W$ packets. The congestion
window is cut in half ($W/2$), and then it increases by one packet
per round-trip time until it reaches $W$ packets again; at which
point, the network drops a packet again and the steady-state model
continues as before. Let us look at a single congestion epoch.

For that simple model, the number of packets sent during the
congestion epoch is
$\frac{w}{2}+(\frac{w}{2}+1)+...(+\frac{w}{2}+\frac{w}{2})=\frac{3w^2}{8}+\frac{3w}{4}$.
For the packet to be lost , both copies sent over the two paths must
be lost. Therefore, the loss rate is
$p=p_1p_2=\frac{1}{number~of~packets}=\frac{1}{\frac{3w^2}{8}+\frac{3w}{4}}
\simeq \frac{8}{3w^2}$ and $W\simeq \sqrt{8/3(p_1p_2)}$. The only
difference from \cite{floyd} is that the round-trip time as
perceived by TCP-over-RAIL is no longer constant, but it depends on
whether a packet is lost on any of the paths. Provided that the
packet is received on at least one path, which has prob.
$(1-p_1p_2$), we are still in the same congestion epoch and
\vspace{-10pt}
\begin{equation}
RTT=
\begin{cases}
RTT_1,~w.p.~(1-p_1)\\
RTT_2,~w.p.~p_1(1-p_2)
\end{cases}
\end{equation}
Therefore, the conditional expectation for RTT is given by
Eq.(\ref{eq:tcp-E[RTT]}); and the TCP throughput over RAIL is on
average:
\begin{equation}
\frac{(number~of~packets)}{ (\frac{W}{2}+1)\cdot E[RTT]}
\simeq...\frac{1.22}{E[RTT]\sqrt{p_1p_2}}
\end{equation}

Essentially, RAIL appears to the TCP flow as a virtual path with
loss rate $p=p_1p_2$ and round-trip time $E[RTT]$. Notice that there
are two factors to take into account in
Eq.(\ref{eq:tcp-rail-formula}): a multiplication in loss ($p_1p_2$)
and an averaging in delay E[RTT]. The loss for RAIL is smaller than
any of the two links: $p>p_1,p>p_2$. The same is not true for the
delay which is a weighted average: $RTT_1<E[RTT]<RTT_2$.

{\em Implications.} Let us now use this simple formula to study the
sensitivity of tcp-over-RAIL throughput to the characteristics of
the underlying paths.

{\bf Fact 1.} {\em TCP throughput is better over RAIL than over any
of the two paths: $T>T_1$ and $T>T_2$. }

{\bf Proof.}
First, consider that $RTT_1=RTT_2=RTT$. Then, the RAIL link is
equivalent to a single link with $p=p_1p_2$, which is better than
any of the two by an order of magnitude. What happens when
$RTT_1<RTT_2$? It is easy to see that RAIL is better than the slower
path (2), because RAIL has both smaller loss and shorter RTT than
the slow path (2): \vspace{-10pt}
\begin{equation}
\frac{T}{T_2}=\frac{1}{\sqrt{p_1}}\frac{RTT_2}{E[RTT]}>1 \cdot 1=1
\end{equation}
Is RAIL better than the faster path (1) as well? RAIL is better in
terms of loss but worse in terms of delay ($E[RTT]>RTT_1$). It turns
out that the multiplicative decrease in loss dominates the averaging
in delay.
\begin{figure}
\begin{center}
\centerline{\includegraphics[scale=0.45]{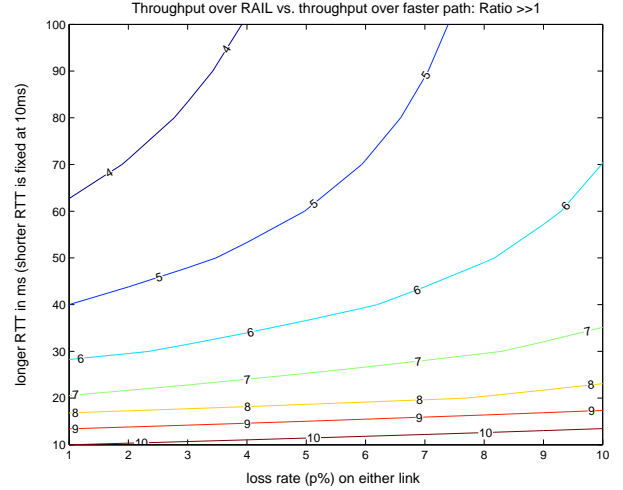}}
\end{center}
\vspace{-20pt} \caption{Consider two paths with the same $p$ and
different $RTT_1<RTT_2$, for the full range of $p$'s and $RTT$'s.
The figure shows the ratio of TCP throughput over RAIL vs. tcp over
the fast link. RAIL performs roughly 10 times better for the range
of practical interest.} \label{fig:tcp-rail-vs-fastlink}
\end{figure}

In Fig.\ref{fig:tcp-rail-vs-fastlink}, we consider $p_1=p_2=p$, we
fix $RTT_1=10ms$ and consider the full range of $p$ and $RTT_2$. We
plot the ratio between the throughput for TCP-over-RAIL vs.
TCP-over-fast-link. \vspace{-10pt}
\begin{equation}
\begin{split}
\frac{T}{T_1}=\frac{1}{\sqrt{p}}\frac{RTT_1}{E[RTT]}~\text{where }~~~~~~~~~~\\
\frac{1}{\sqrt{p}}>1~~~\text{and}~
\frac{RTT_1}{E[RTT]}=...=\frac{1+p}{1+p\frac{RTT_2}{RTT_1}} \le 1
\end{split}
\end{equation}

We see that tcp does 4-10 times better over RAIL than over the fast
link (1), for all practical cases: loss rates up to 10\% and
difference in delay up to 100ms. Indeed, the difference in $RTT$
cannot be exceed some tens of milliseconds (e.g. due to propagation
or transmission ), and $p$ should be really small, except for short
time periods.

{\em How many paths?} For n paths with characteristics $(p_i,
RTT_i), i=1..n$, where $RTT_1<RTT_2<...<RTT_n$, and following
similar derivations, we find that:
\begin{equation}
\begin{split}
T(n)=\frac{1.22}{E[RTT]\sqrt{p_1p_2...p_n}}, \text{where:}~~~~~~~~~~~~\\
E[RTT]=\frac{[RTT_1+RTT_2p+...RTT_np^{n-1}](1-p)}{1-p_1...p_n}\\
\end{split}
\label{eq:tcp-kpaths}
\end{equation}
The multiplicative factor $\sqrt{p_1..p_k}$ dominates the averaging
E[RTT]. Also large RTTs have discounted contributions. For
$p_1=p_2=...p_n$, $T(n)$ is a convex increasing function of $n$,
which implies that adding more paths of similar loss rate, improves
throughput but with decreasing increments.

\subsubsection{\label{sec:reordering}Testbed Experiments on Reordering and TCP}

In section \ref{sec:reordering-general}, we saw that RAIL does not
introduce reordering if both paths are well behaving, but may
convert loss on the fast path to late - and at the extreme even
out-of-order packets under some conditions ($dt\le d_2-d_1$). It is
well known that  reordering may have a reverse effect on TCP, as it
falsely triggers the fast retransmit. In this section, we use
testbed experiments to show that, even in cases that RAIL converts
loss to reordering, this is actually beneficial for TCP.

\begin{figure}
\begin{center}
\centerline{\includegraphics[scale=0.35,angle=-90]{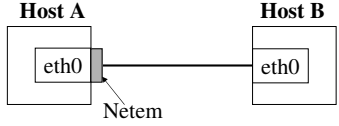}}
\end{center}
\vspace{-10pt} \caption{Simplified experimental setup for testing
the effect of reordering vs. loss on TCP.}
\label{fig:testbed-Single} \vspace{-10pt}
\end{figure}

\begin{figure}
\begin{center}
{\includegraphics[scale=0.4]{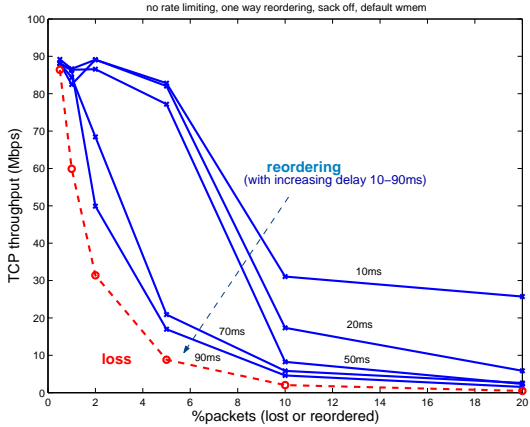}}
\end{center}
\vspace{-10pt} \caption{\label{fig:loss-vs-reorder} Testbed
experiments comparing the effect of loss vs. reordering on the
throughput of a single TCP flow.} \vspace{-10pt}
\end{figure}

Recall that RAIL does not cause reordering, it only translates loss
to reordering. Therefore, the fair question to ask is not how ``TCP
does with reordering vs. without reordering'' but instead ``how TCP
does with  $x\%$ of packets arriving out-of-order vs.  $x\%$ of
packets being lost''.

{\em{\bf Fact 3-revisited.} Better late than never (and the earlier
the better)}. We used the simplified testbed shown in
Fig.\ref{fig:testbed-Single} to inject a controlled amount of loss
and reordering, using Netem, on a single TCP flow.
Fig.\ref{fig:loss-vs-reorder} shows the results of the comparison.
First, we introduced x\% of loss, ranging from 0 to 20\%; the TCP
throughput is shown in dashed line.  Then we introduced x\% of
reordering for a range of reordering gaps/delays, i.e. the packets
arrive 10-90ms later than they should; the resulting TCP throughput
is shown in a separate bold line for each delay value. We see that
TCP performs much better with reordering than with loss, therefore
it is  indeed better to receive packets ``late than never''. Not
surprisingly, the less the delay in delivery, the better the
performance.

Furthermore, TCP has today several default options to deal with
reordering: including SACK, DSACK and timestamps. We found that
turning SACK on further improved the performance of TCP under
reordering in Fig.\ref{fig:loss-vs-reorder}. In summary, we expect
RAIL to help TCP for all practical cases, i.e. for small loss rates
and delay differences between the paths in the order of 10-50ms. As
an extreme measure, one can use the delay padding mechanism not only
for voice, but also as a TCP ordering buffer to completely eliminate
reordering.

\section{Future Directions}
\label{sec:discussion}

We envision a RAIL-network architecture, where RAILedges are control
points that use path redundancy, route control and
application-specific mechanisms, to improve WAN performance.

\begin{figure}
\centerline{{\includegraphics[scale=0.28,angle=-90]{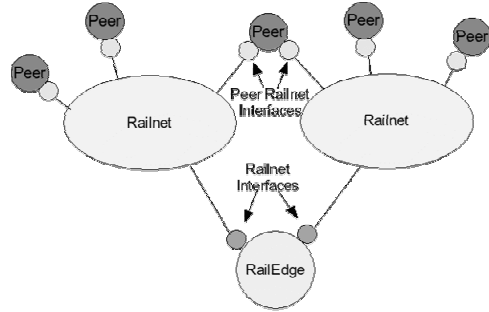}}}
\vspace{-10pt} \caption{RAILnet: a virtual multipoint reliable
network} \label{fig:railnet} \vspace{-10pt}
\end{figure}

A first extension has to do with topology. So far, we considered two
RAILedge devices connecting two remote sites via multiple redundant
links. We envision that this can be generalized to a virtual
multipoint network or {\em RAILnet}, where multiple edge networks
are reliably interconnected to each other, as shown in
Fig.\ref{fig:railnet}. Each participating edge network is located
behind its own RAILedge, and each RAILedge pair communicates over at
least two Internet links. The Railnet interface represents the local
point of attachment to a Railnet and should present itself as a
regular interface to a multi-access subnet.

Second, we are interested in combining the proactive replication of
RAIL with some kind of route control, in particular (i) selection of
the right subset of physical paths within the same RAIL and also
(ii) dynamically switching among them. In this paper, we focused on
the first part (i.e. at combinations of paths with various
characteristics, at different number of paths, at paths that are
similar or different from each other) and tried to give
recommendations on how to statically select among them. The second
aspect is dynamic switching among sets of paths. We expect this to
be a less constrained than single-path switching, because (i)
redundant transmission is robust to short-lived problems and (ii)
physical paths tend to have consistent behavior in the long time
scales. Therefore, RAIL should relieve much of the urgency in
dynamic path switching decisions.

One could further enhance the functionality of RAILedge. So far, we
focused on replication of packets over multiple paths. Several other
functions can be naturally added on an edge network device,
including monitoring and path switching, compression,
quality-of-service mechanisms, protocol specific acceleration. For
example, one could decide to RAIL part of the traffic (e.g. VoIP or
critical applications) and use striping for the remaining traffic;
this could correspond to RAIL-0 in the raid taxonomy \cite{raid}.

There are some additional interesting questions, we are currently
pursuing as a direct extension of this work. First, we continue to
study TCP over RAIL, using more accurate TCP models, and considering
also short-lived connections;
 we are also working on a modification of our delay-padding
algorithm, to remove reordering at the receiving RAILedge. Second,
we are investigating the effect of RAIL on the rest of the traffic.
E.g. when there is significant disparity in bandwidth, we expect
RAIL-ed TCP to cause congestion on the limited-bandwidth path.
Furthermore, what is the interaction between competing RAILs?
Finally, it would be interesting to explore the benefit of adding
additional RAILedges in the middle of the network.

The RAILnet architecture can be incrementally deployed by gradually
adding more RAILedges. If widely deployed, it has the potential to
fundamentally change the dynamics and economics of wide-area
networks.

\vspace{-20pt}
\section{Conclusion}
\label{sec:conclusion} \vspace{-10pt} We proposed and evaluated the
Redundant Array of Internet Links (RAIL) - a mechanism for improving
packet delivery by proactively replicating packets over multiple
Internet Links. We showed that RAIL significantly improves the
performance in terms of network- as well as application-level
metrics. We studied different combinations of underlying paths: we
found that most of the benefit comes from two paths of carefully
managed; we also designed a delay padding algorithm to hide
significant disparities among paths. RAIL can be gracefully combined
with and greatly enhance other techniques currently used in overlay
networks, such as dynamic path switching. Ultimately, it has the
potential to greatly affect the dynamics and economics of wide-area
networks.

\vspace{-20pt}
 {\footnotesize

}

\end{document}